\documentclass{article}

\usepackage{graphicx}%

\usepackage{tabularx}
\usepackage{url}

\usepackage[numbers,sort&compress]{natbib}

\usepackage{authblk}

\usepackage[margin=1.2in]{geometry}

\usepackage[colorlinks,citecolor=blue,urlcolor=blue]{hyperref}
\usepackage{doi}

\title{Overcoming Barriers to Computational Reproducibility}

\author[1,2]{Roman Hornung}
\author[3,4,*]{L\'{a}szl\'{o} N\'{e}meth}
\author[5]{Oleksandr Zadorozhnyi}
\author[6]{Theresa Ullmann}
\author[6,7]{Michael Kammer}
\author[8,9]{Rebecca Killick}
\author[10]{Christopher J.\ Paciorek}
\author[11]{Julien Chiquet}
\author[2,12]{Moritz Herrmann}
\author[13,14]{Lucija Batinovi\'{c}}
\author[14]{Rickard Carlsson}
\author[15]{Pierre Neuvial}
\author[16]{Boris Hejblum}
\author[17]{Julia Wrobel}
\author[12,2]{Anne-Laure Boulesteix}
\author[3]{Karsten Tabelow}

\affil[1]{Department of Statistics, Ludwig-Maximilians-Universit\"at (LMU), Munich, Germany}
\affil[2]{Munich Center for Machine Learning (MCML), Munich, Germany}
\affil[3]{Weierstrass Institute for Applied Analysis and Stochastics, Berlin, Germany}
\affil[4]{Max Planck Institute for Demographic Research, Rostock, Germany}
\affil[5]{Department of Computer Science, TUM School of Computation, Information and Technology, Technical University of Munich, Munich, Germany}
\affil[6]{Institute of Clinical Biometrics, Center for Medical Data Science, Medical University of Vienna, Vienna, Austria}
\affil[7]{Division of Nephrology and Dialysis, Department of Medicine III, Medical University of Vienna, Vienna, Austria}
\affil[8]{School of Mathematical Sciences, Lancaster University, Lancaster, United Kingdom}
\affil[9]{School of Mathematical and Statistical Sciences, Clemson University, Clemson, USA}
\affil[10]{Department of Statistics, University of California, Berkeley, USA}
\affil[11]{UMR MIA Paris-Saclay, INRAE, AgroParisTech, Universit\'{e} Paris-Saclay, Palaiseau, France}
\affil[12]{Institute for Medical Information Processing, Biometry, and Epidemiology, Faculty of Medicine, Ludwig-Maximilians-Universit\"at (LMU), Munich, Germany}
\affil[13]{Department of Behavioural Sciences and Learning, Linköping University, Linköping, Sweden}
\affil[14]{Department of Psychology, Linnaeus University, V\"axj\"o, Sweden}
\affil[15]{Institut de Math\'{e}matiques de Toulouse (IMT), CNRS, Universit\'{e} de Toulouse, Toulouse, France}
\affil[16]{SISTM, U1219 Bordeaux Population Health, Universit\'{e} de Bordeaux / INSERM / Inria, Bordeaux, France}
\affil[17]{Department of Biostatistics and Bioinformatics, Emory University, Atlanta, USA}
\affil[*]{Corresponding author: L\'{a}szl\'{o} N\'{e}meth, nemeth@wias-berlin.de}

\begin{document}

\maketitle

\begin{abstract}
\noindent Computational reproducibility, the possibility for independent researchers to exactly reproduce published empirical results, is fundamental to science. Despite its importance, the proportion of research articles aiming for reproducibility remains low and uneven across disciplines. Barriers include a perceived lack of incentives for researchers and journals, practical challenges in preparing reproducible materials, and the absence of harmonised standards of reproducibility processes and requirements by journals. Existing guidance is often highly technical, reaching mainly those already engaged with reproducible research.  In this paper, we first synthesize evidence on the benefits of reproducibility for both authors and journals. Drawing on our extensive experience in reproducibility checking at various journals, we then put forward concise, pragmatic guidelines for creating reproducible analyses across disciplines. We further review current reproducibility policies of selected journals, illustrating the substantial heterogeneity in requirements and procedures. Motivated by the latter, we propose conceptual foundations for a harmonised multi-tier system of reproducibility standards that could support transparent, consistent assessment across journals and research communities. Our goal as journal (reproducibility) editors and contributors to the MaRDI initiative is to encourage broader adoption of reproducibility practices, in particular by lowering practical barriers for authors and journals.
\end{abstract}

\section{Introduction}

Computational reproducibility is a fundamental principle of scientific rigour and credibility. In this paper, we define computational reproducibility as the possibility for independent researchers to exactly reproduce the final results of an analysis without any involvement of its original authors. Note that terminology in this area is not entirely consistent across fields: in some disciplines, what we refer to as computational reproducibility is instead termed replicability. In contrast, according to the terminology proposed by the U.S.\ National Academies of Sciences, Engineering, and Medicine \cite{NationalAcademies:2019}, replicability refers to conducting a new study with newly collected data and obtaining findings that are consistent with the previous study.

Ensuring computational reproducibility requires not only the provision of the analyzed data, but also the complete programming code used for the analysis (or, in the case of point-and-click software, the corresponding analysis files or workflows), including all specifications of the software environment. The importance of computational reproducibility extends across all forms of data use, including fully open and restricted data. Increasingly, many datasets are prepared according to the FAIR principles---findable, accessible, interoperable, and reusable \cite{Wilkinson:2016}---which provide guidance on metadata standards and on enabling the effective use and reuse of data. Practical possibilities for independent verification naturally differ across levels of data accessibility.

Throughout this paper, we focus on computational reproducibility in studies that analyze real data, simulated data, or a combination of both, irrespective of discipline. Our focus therefore encompasses all research fields in which results depend on computational analyses of real or simulated data. This excludes forms of reproducibility that involve laboratory, field, or qualitative research settings.

In a pioneering work, Claerbout and Karrenbach \cite{Claerbout:1992} highlighted the value of ensuring reproducibility by linking publications to the underlying analyses through electronic documents. Buckheit and Donoho \cite{Buckheit:1995} later argued that scientific articles presenting the results of data-based analyses are merely advertisements for the underlying research, whereas the code and data constitute the actual scientific work. Indeed, a scientific article involving computational research could not exist without the code and data that generated its results. While the article is indispensable for communicating results to the scientific community, the validity of the findings relies on the underlying code and data, not the manuscript itself.

However, the concept proposed by Buckheit and Donoho \cite{Buckheit:1995} does not, in itself, explain why code and data should be made publicly available. A justification for publishing them can be drawn from an analogy to pure mathematics, where research findings take the form of statements established and communicated by proofs, and where it would be inconceivable to withhold these proofs from the public. In empirical research, results, such as figures or tables, analogously should be supported by available code and data, thus open to verification. It is in the scientific community’s interest and increasingly its expectation that published research is verifiable and that statements, claims, and results are valid and rigorously checked, so that further research can build upon already discovered facts, solutions, and outcomes with certainty. Beyond this fundamental role as the empirical foundation---or \lq\lq proof''---of research results, the publication of code and data offers many benefits for both researchers and journals, which we discuss in Section~\ref{sec:benefits}.

In recent decades, the proportion of empirical research articles (i.e.\ articles based on the analysis of data) that aim for reproducibility by providing the necessary materials has increased in some disciplines but not in others, and this proportion remains low when considering the scientific literature as a whole \cite{Hardwicke:2024, Sanchez:2024, Serghiou:2021, Hamilton:2023}. Moreover, even when reproducibility materials are provided, the corresponding published results are frequently still not reproducible \cite{Samuel:2024}. In certain data-intensive fields, such as empirical machine learning, practices have moved closer to what Donoho \cite{Donoho:2024} refers to as \lq\lq frictionless reproducibility''---workflows in which code, data, and computational environments are organised so that analyses can be reproduced with minimal effort (supported, e.g., by platforms such as OpenML \cite{Vanschoren:2013}). However, such communities remain the exception rather than the rule. In addition to the fact that, in many cases, the underlying data cannot be made publicly available---whether due to legal, ethical, or FAIR-compliant access restrictions---we identify three main factors contributing to this persistently low share of reproducible studies.

The first factor, a lack of incentives to ensure reproducibility, concerns both individual researchers and journals. For researchers, making analyses fully reproducible requires additional time and effort, and in the current academic reward system this work is frequently insufficiently recognised \cite{Smaldino:2016}. Therefore, researchers often perceive little direct personal benefit in publishing their code and data, although there are various self-serving advantages to doing so. Numerous scientists have called for journals to make the publication of code and data supplements mandatory for all empirical studies \cite{LeVeque:2012, Stodden:2013, Goldacre:2019, DeBlanc:2020, Sanchez:2021, Trisovic:2022}, and the importance of more systematic assessment of computational reproducibility has been highlighted \cite{Lindsay:2023}. However, journals often fail to recognise the self-serving benefits associated with requiring publication of such materials. Consequently, only some journals provide guidance on supplementary code and data preparation and publication (as discussed in Section~\ref{sec:journals}).

The second factor is that many researchers feel overwhelmed by the practical demands of ensuring reproducibility. Greater engagement with the topic would likely alleviate this perception, but most researchers currently devote little attention to reproducibility compared to writing their scientific articles. Although the existing literature on this topic is generally of high quality, it tends to be very detailed and often inaccessible to readers without prior knowledge. Consequently, this literature primarily reaches those already familiar with, or positively inclined toward, the subject. Researchers with little prior exposure to reproducibility are unlikely to consult, let alone follow, lengthy technical recommendations on writing clean code or creating reproducible analyses. Our experience with reproducibility reviews in various journal contexts supports this observation: while some journals increasingly receive submissions whose initial materials already meet basic reproducibility criteria, it remains common that authors encounter difficulties in satisfying reproducibility requirements, even when structured guidance or checklists are provided.

The third factor, the insufficient institutionalisation of reproducibility, concerns the role of journals and other actors in scientific publishing. Among the journals that have already introduced measures to promote reproducibility, requirements and procedures vary considerably. This heterogeneity makes it difficult to implement reproducible research practices consistently across journals and leads to similar works being evaluated very differently depending on where they are published. Furthermore, overly complex or poorly coordinated procedures make code checking more labor-intensive for journals, whereas vague guidelines have little effect in practice. Overall, there is a lack of coordinated and practical structures at the system level that would anchor reproducibility as an integral part of the publication process. Broader problems of metric-driven and fraudulent publishing in the mathematical sciences are discussed by  Agricola \emph{et al.}\ \cite{Agricola:2025}. It is also noteworthy to mention the positive contribution of designated reproducibility editors for journals.

In this paper, we contribute by addressing the three factors outlined above through a synthesis of existing evidence and through both practical guidance and conceptual suggestions. Our aim is to lower barriers to reproducibility for individual researchers while also outlining ways in which journals and institutions can support more consistent and pragmatic practices. The paper is intended primarily for authors and journal editors, and we aim to make our recommendations accessible to readers with varying levels of computational expertise, as even experienced researchers can encounter challenges when creating fully reproducible analyses.

First, we summarize the benefits of reproducibility for individual researchers and journals, as documented in the literature. Second, we provide concise, easy-to-follow instructions for authors to ensure reproducibility. These are informed by our experience reviewing code and data supplements of many hundreds of articles, which has revealed common pitfalls and recurring difficulties for authors. This experience reflects our collective involvement as (reproducibility) editors for several journals with established reproducibility procedures, namely the Biometrical Journal, Computo, Journal of Statistical Software, Journal of the American Statistical Association, and Meta-Psychology. The aim of these instructions is to encourage broader adoption of reproducible practices by making them accessible even to those with little prior experience or initial motivation in ensuring reproducibility.

Third, we discuss how journals and institutions can promote reproducibility through clearer and realistically implementable framework conditions. We summarize current reproducibility practices across several journals, which highlights the substantial heterogeneity in existing guidelines. Building on this overview, we outline conceptual ideas for more consistent and structured reproducibility standards, informed in part by our experience as contributors to the Mathematical Research Data Initiative (MaRDI). Such standards could help journals communicate expectations more transparently, and support more uniform assessment of reproducibility across studies and disciplines.  We present them as starting points for future developments in scientific publishing rather than prescriptive standards.

\section{Benefits of reproducible research for authors and journals}
\label{sec:benefits}

Researchers and journals tend to perceive only limited self-serving benefits from adopting reproducible practices, which likely reduces motivation to invest effort in this area. If the perception of researchers' incentives were strengthened, the proportion of reproducible results could increase substantially, even in the absence of strict journal policies. Likewise, if journals more clearly recognised the advantages of implementing reproducibility policies, broader adoption of such policies would likely follow, further promoting reproducible research. Importantly, such policies need not involve full reproducibility checks; they span a wide spectrum, ranging from simple expectations, such as requiring that code and, when possible, data be made available, to more extensive review procedures. Even lightweight measures can improve transparency and contribute meaningfully to reproducible research.

In the following subsections, we describe advantages for both authors and journals, many of which apply to both groups.

\subsection{Benefits for researchers' own workflow and future work}

For individual researchers, ensuring reproducibility offers several practical advantages for their own future work, such as easier reuse of (clearer) code, greater confidence in the correctness of results, and mitigation of research waste. Researchers are typically the primary future users of their code, and thus benefit most from its availability and from the results' reproducibility. The prospect of sharing code publicly increases attention to code quality \cite{Assel:2018}, which in turn facilitates later navigation and modification of analyses, for example during manuscript revision or when extending the work in follow-up projects \cite{Alston:2021}. Given the considerable time that often elapses between initial analysis and publication, the documented versions of add-on software packages are crucial. Proper documentation ensures exact reproducibility should the analysis be modified later, where package functionalities may have changed. Clean, well-documented code benefits other members of the research team \cite{Rasmussen:2023}, streamlining further analyses and collaboration on the same codebase. Researchers who consistently publish the code and data of their projects will also have secure access to them in the future, regardless of whether these researchers change employers or computers \cite{Gomes:2022}.

Making the code and data publicly available may also reduce the perceived psychological burden on some authors \cite{Vable:2021}. Although authors retain full responsibility for the integrity of their work, sharing allows reviewers to inspect the code, which some authors may perceive as a form of reassurance and support \cite{Goldacre:2019, Sanchez:2021}. Conversely, others may worry about judgments on code quality and potential errors being identified, meaning this psychological effect is likely to vary across individuals.

Sharing code and data signals confidence in one's work and can attract collaborators who build on the shared materials in transparent and mutually beneficial ways. Even when others reuse the code independently, the original authors remain part of the foundation of that work and are typically acknowledged, reinforcing rather than undermining their contribution. By contrast, some researchers worry that others might use their code and data to perform analyses or develop extensions they themselves planned. Yet such concerns rarely materialize in practice: the risk is low, because the original authors are most familiar with the analysis and retain deep, specific knowledge of the research context \cite{Gomes:2022}.

\subsection{Scientific impact and career benefits}

Reproducibility also offers several advantages for scientific quality and researchers' careers. When authors pay more attention to code quality and review their materials prior to publication, the likelihood of detecting errors before the article appears increases substantially \cite{Gomes:2022}. Conversely, later retractions issued because of analytical errors can disrupt the iterative process of knowledge building and may create severe challenges for both authors and journals \cite{Sanchez:2021, Goldacre:2019}. Reproducibility policies can plausibly help lower the risk of retractions. By promoting code transparency and, in some cases, through pre-publication reproducibility checks conducted by journals, such policies increase the likelihood that analytical errors are detected before publication, even though the level of reproducibility checks that journals can realistically provide will vary with their resources.

Methodological articles often propose complex algorithms. If the implementation code is not available, these algorithms are difficult to apply in practice, which can greatly reduce their impact and the number of citations \cite{Celi:2019}. Moreover, their inclusion in comparison studies becomes substantially more difficult, which further limits their visibility. In this context, authors should not be expected to reimplement methods for which no usable code exists; such reimplementations can be error-prone due to insufficiently detailed methodological descriptions in the original work and place an unreasonable burden on those conducting comparison studies.

Readily available code increases the likelihood that other research groups build on or extend a published algorithm, thereby amplifying its impact within the methodological community. Moreover, a clean implementation is more easily adopted and built upon than an approach that is only theoretically described, even if promising---especially when implementation details are missing or the method is complex \cite{Benureau:2018}.

Individuals who are primarily interested in applying proposed approaches often focus more on the implementations than on the paper's theoretical descriptions. These individuals frequently contact the authors with questions about the application. Such inquiries provide valuable insights into common difficulties in practice and can lead to the correction of errors and improvements in methodology \cite{Benureau:2018}.

An important reason why publishing the code and data can result in more citations is that it has become common practice to cite code that is reused---not only within the computational field, but especially in applied disciplines, where researchers tend to depend on available implementations because implementing the underlying algorithms can be challenging. Also, citation counts tend to be higher when supplements are uploaded to a repository that assigns DOI numbers \cite{Goldacre:2019}. Overall, it has repeatedly been shown that papers making their code \cite{Vandewalle:2012, Maitner:2024} and/or data \cite{Gomes:2022, Piwowar:2007, Colavizza:2020} publicly available receive more citations. H\"{o}ffler \cite{Hoefler:2017} also showed for economics journals that articles published in journals requiring reproducibility are cited statistically significantly more often, accounting for differences between journals and the general time trend.

Implementing proposed algorithms in software packages on established platforms (such as GitHub, CRAN for R, or PyPI for Python) further increases their practical applicability compared to providing their implementations solely in code and data supplements to a methodological paper. It is therefore likely that package-based implementations also have a positive effect on citation counts. In a related context, Jalili \emph{et al.}\ \cite{Jalili:2020} showed that the citation count of biomedical software packages increased substantially after their inclusion in package management systems.

Funding agencies are giving reproducibility an increasingly important role by requiring open data and code, mandating data management plans, and in some cases even covering long-term data storage costs \cite{Gomes:2022, Streiber:2025}. Authors can therefore improve their chances of obtaining funding by consistently making their analyses reproducible and highlighting this in their funding proposals.

Publishing reproducible results also increases the credibility and trustworthiness of researchers \cite{McKiernan:2016}. This signals scientific diligence and transparency to the community and can influence the peer-review process beneficially \cite{Alston:2021}.  Occasionally, reviewers may run the shared code themselves and form critical impressions when legitimate issues arise. From a scientific perspective, such cases are valuable, as they help strengthen the quality of published work. They, however, appear to be relatively rare in practice and should not discourage authors from sharing their materials, especially given the substantial credibility benefits of transparency for well-executed work.

\subsection{Specific benefits for journals}

The advantages of reproducibility policies for journals and the broader research ecosystem are significant, as discussed for example by Stodden \emph{et al.}\ \cite{Stodden:2013}. Primarily, these policies allow journals to function as a quality mark signalling their commitment to scientific rigour and strengthening their reputation. Such policies can also provide incentives for authors, who may benefit from publishing in journals that visibly endorse rigorous and transparent research practices. Indeed, Stodden \emph{et al.}\ \cite{Stodden:2013} found that journals with a high impact factor are more likely to adopt such policies, although the direction of causality remains unclear.

This increased attractiveness to submitting authors may allow journals to be in a stronger position to demand reproducibility---even though this potential is not necessarily recognised by all of them. Conversely, journals that impose such quality requirements may appear more selective to authors, which can increase their desirability even if their impact factor is not particularly high.

A clear reproducibility policy---even one that only requires authors to submit code and, when possible, data without performing reproducibility checks---offers dual operational benefits for journals. First, it acts as a quality filter, improving submissions by discouraging authors who place less emphasis on rigour or transparency. Second, clearly specifying requirements reduces the chance of receiving submissions in an unacceptable form.

\section{Straightforward guidelines for creating reproducible analyses}
\label{sec:guidelines}

\subsection{General considerations}

As mentioned in the introduction, one of our aims is to provide concise, easy-to-follow guidelines that keep the effort required to implement them low. Readers seeking more detailed guidance on good coding practices, advanced techniques, or technical intricacies are referred to the existing literature (e.g. \cite{Sandve:2013}, \cite{Wilson:2017}, \cite{Sanchez:2021}). A set of minimum recommendations focused primarily on data can be found in \cite{Jenkins:2023}. Our experience shows that even seemingly obvious aspects of reproducible research are often overlooked in practice. Therefore, we decided to include them here to help prevent common pitfalls. Note also that many of the aspects are not strictly required for ensuring reproducibility, but they make the process considerably easier for those attempting to reproduce the results. While more sophisticated solutions exist for many of these aspects, the recommendations presented here already provide a solid foundation for creating reproducible analyses in a pragmatic and accessible way.

Our recommendations provide a general foundation that is applicable across disciplines. When available, subject-specific standards (e.g., PSYCH-DS for psychology) and local or national guidelines can offer additional, domain-tailored guidance. These resources may further support authors in organising their materials according to conventions within their research community.

It is highly beneficial to apply the following tips from the very beginning of a research project, as many of them are considerably more time-consuming to implement retrospectively. Crucially, sharing imperfect code and data is still better than not sharing it at all, even if not all recommendations have been followed. 
We provide R- and Python-specific hints as well, but the recommendations can be applied in other programming languages in a similar way.

When preparing the code and data supplement, it can be helpful to occasionally adopt the perspective of someone unfamiliar with the underlying study. This makes it easier to identify parts of the material that might be unclear or difficult to use without detailed knowledge of the study.
Although these guidelines are primarily designed for authors preparing code and data supplements, they can also serve as a helpful reference for reviewers or reproducibility editors.

\begin{figure}[htbp]
  \centering
  \includegraphics[width=\textwidth]{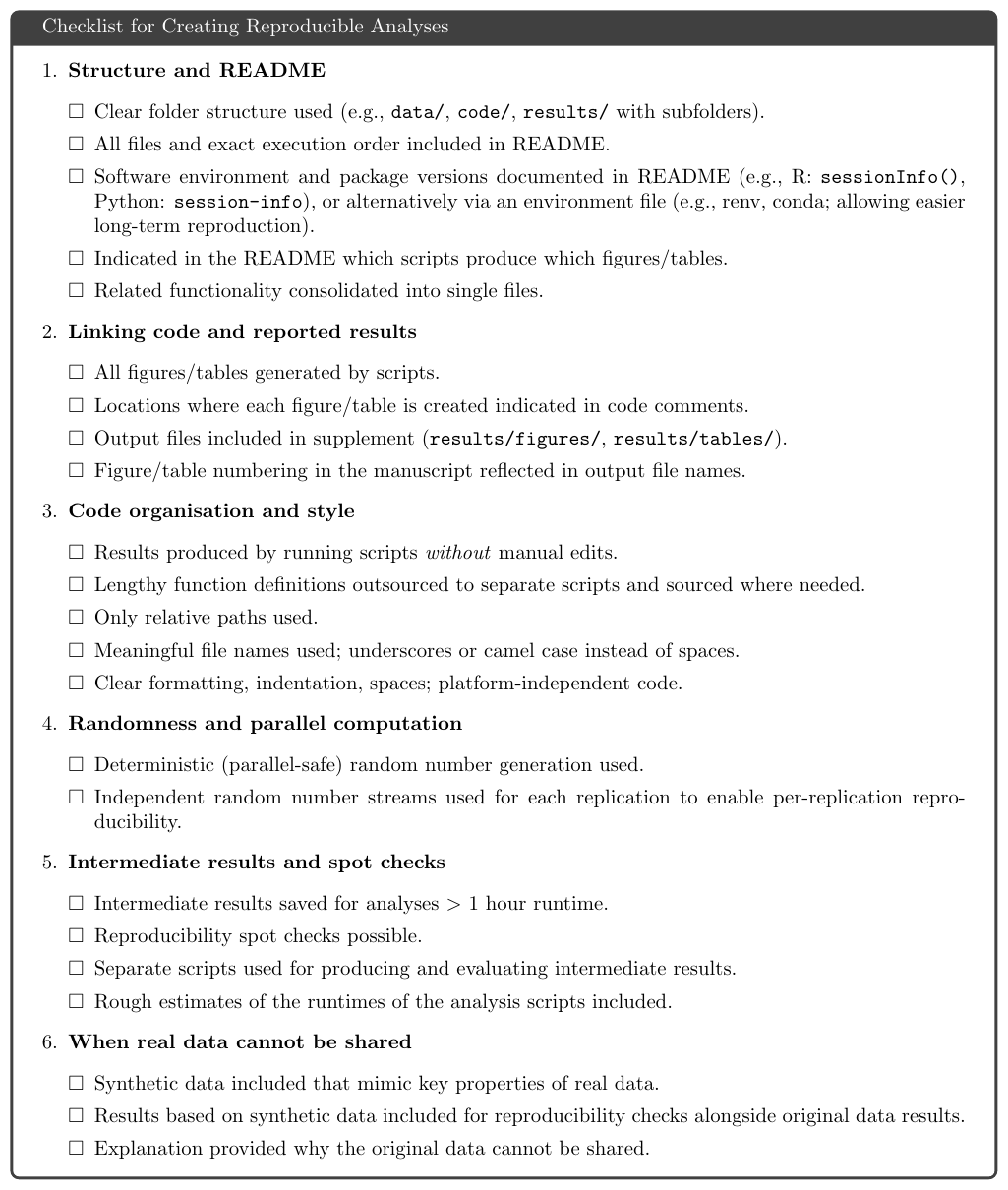}
  \caption{Checklist for creating reproducible analyses.}
  \label{fig:checklist}
\end{figure}

Before describing the guidelines in detail, we provide in Figure~\ref{fig:checklist} a checklist as a concise summary of the main recommendations. Note that not all points in this checklist may be fully understandable without reading the detailed guidelines. The checklist primarily serves as a quick first reference for identifying which aspects may warrant closer reading in the detailed guidance below, as the relevance of the recommendations to individual readers depends on their prior familiarity with reproducible research practices. We also provide a one-page PDF version of the checklist in the supplementary material, formatted as a print-friendly version.

\subsection{Structure and README}

The code and data supplement should contain all materials necessary to independently reproduce the results presented in the associated manuscript. This typically includes the complete analytical code, the relevant dataset or datasets (where possible) together with accompanying codebooks (files that explain all variables in the dataset(s)), and a README file.
The code and data supplement should follow a simple folder structure with clear directory names such as \texttt{data}, \texttt{code}, and \texttt{results}, the latter including subfolders like \texttt{intermediate} (for intermediate results, see below), \texttt{figures}, and \texttt{tables}.
To maintain a clear and manageable file structure, related functionality should be consolidated into a single file rather than spread across many files that serve the same purpose.

A README file is of central importance. It should give a brief overview of all files in the code and data supplement and specify the order in which scripts must be executed to reproduce the final results (figures and tables) presented in the article and supplement. Complex descriptions of the analysis design should be avoided, as they may overwhelm readers unfamiliar with the study details. Instead, simply indicate which figures and tables each script produces, referencing their respective numbers (e.g., Figure 1, Table 1).

For long-term reproducibility, the software environment must also be clearly documented, including the versions of all additional packages used. In R, this can be easily achieved by loading all relevant packages into the workspace and copying the output of \texttt{sessionInfo()} into the README file. In Python the \texttt{session-info} package produces similar output with the information on the workspace environment. Alternatively, long-term reproducibility can be further supported by providing an environment specification file (e.g., via renv in R or conda in Python), which enables automatic restoration of the software environment.

A codebook containing the variable names, variable descriptions, variable types (e.g., numeric), and possible values the variables can take is not strictly required for computational reproducibility. However, it is strongly recommended, as it greatly enhances transparency, facilitates understanding and reuse of the data, and aligns with established open-data principles. This information also allows checking for potential errors.

An illustrative example of an appropriate structure for the code and data supplement and the README file, with additional details, is provided in Section A of the supplementary material.

\subsection{Linking code and reported results}

All figures and tables in the article and supplement that contain empirical results should be reproducible.
Comments should be used to clarify the purpose of individual code sections and, specifically, to clearly indicate where each figure or table is generated, citing the corresponding numbers of these figures and tables.

The code should produce all figures and tables, and these output files should also be included in the code and data supplement. The structure of the saved tables (row and column order plus any rounding of numbers) should match the corresponding tables in the article; this makes it easier to compare the reproduced tables with those in the manuscript. Ideally, file names should correspond to those used in the article (e.g., figure\_1.pdf, table\_1.csv).

\subsection{Code organisation and style}

Fundamentally, final results should be obtainable by running the corresponding scripts, without requiring any manual changes to the code.
Analysis scripts producing the results should avoid lengthy function definitions unless the project is simple; for anything beyond basic analyses, such functions should be collected in separate scripts and sourced in the analysis scripts to improve clarity. For key self-written functions, input and output parameters should be briefly documented.

Within files, only relative paths, defined with respect to the root folder, should be used.  The files themselves should all have meaningful names, avoiding spaces and using underscores or camel case instead.

Code should be easy to read, with spaces after commas and around operators, and proper indentation. Such formatting can be performed automatically using linters or other code-formatting tools. In R and Python, libraries such as \texttt{formatR} and \texttt{Ruff}, respectively, provide these features.

The code should make as few assumptions about the execution environment as possible. In particular, it should be platform-independent whenever feasible (unless strong reasons exist, such as the use of fork-based parallelization) and software-independent (e.g., avoid relying on IDE-specific APIs that require R code to be executed from within RStudio).

Finally, the code should aim to remain as simple as feasible and, wherever possible, rely only on a small number of widely used packages. This reduces the need to install dependencies, minimizes potential conflicts and promotes long-term reproducibility.

\subsection{Random seeds and parallel computations}

For analyses that include random elements (such as statistical simulations), it is essential to set the seed of the random number generator to an arbitrary, but fixed, value to ensure exact reproducibility. In R, for example, this can be done with the command \texttt{set.seed(1234)}, or in Python using \texttt{numpy.random.default\_rng(1234)}.

In parallel computations, however, setting a single global seed is usually not sufficient, as parallel workers may generate their own random number streams. To obtain reproducible results, it is advisable to rely on software mechanisms that explicitly create parallel-safe, deterministic random number streams for each replication. Many modern environments provide such functionality, for example the \texttt{future} ecosystem in R (via seed-handling options in packages such as \texttt{future.apply} or \texttt{foreach} combined with \texttt{doRNG}) and NumPy's SeedSequence-based random number generation in Python. These tools are designed to reduce the risk of overlapping streams and to make correct use of random numbers in parallel settings easier.

Using such mechanisms ensures that each replication is associated with its own deterministic random number stream. As a consequence, results do not depend on the number of available CPU cores, analyses can be reproduced on different systems or with different parallelization backends, and individual replications can also be recomputed sequentially if needed. Importantly, this makes it straightforward to check the reproducibility of selected replications, as explained in the next subsection. A simple R and Python example demonstrating how to implement such parallel-safe random number streams and how to reproduce individual replications is provided in Section B of the supplementary material.

\subsection{Intermediate results and reproducibility spot checks}

For computationally intensive analyses that take more than about an hour to run, it is helpful to save intermediate results. In simulation studies, these consist of the raw results obtained for each simulated data set. While intermediate results are not strictly required for reproducibility, they make the process more flexible and the results easier to reproduce. Approximate runtimes should be provided for each analysis script, together with a brief note on the hardware used (e.g., ``$\sim 5$ minutes on a standard laptop'').

Saving intermediate results serves two main purposes.  First, tables and figures can be generated quickly from the stored intermediate results without re-running the entire analysis. This benefits both the person reproducing the results and the original code author, for instance, if adjustments to the presentation are later desired.
Second, with appropriate handling of random number generation, particularly in analyses that use parallel computation, intermediate results enable so-called reproducibility spot checks. In such checks, only a few arbitrarily selected partial results (e.g., individual replications in simulation studies) are recomputed and compared with the corresponding stored results. If these are identical, the spot check is successful, and it can be concluded that the full analysis is likely reproducible.

A straightforward way to ensure that individual partial results can be reproduced is to use reproducible random number streams for each replication (see previous subsection). This allows anyone reproducing the analysis to rerun selected replications and compare the newly obtained results with the stored ones. Beyond reproducibility spot checks, this approach also makes it possible to inspect individual unusual or outlying results in more detail to investigate the reasons behind such anomalies.

The code that generates the intermediate results should be easy to identify and run on its own. This makes it straightforward to reproduce individual intermediate results when needed.

An alternative, easy-to-implement strategy for making computationally intensive analyses easier to reproduce is to rerun the analysis with a much smaller number of replications (e.g., 10 instead of 1,000) and provide these corresponding results in the code and data supplement in addition to the full results. Anyone checking reproducibility can then use the smaller number of replications to verify that they obtain the same results as those included in the supplement. Crucially, the smaller run should use the same code as the larger run, simply with the number of replications reduced.  Then if the results match, this provides strong evidence that the full analysis is reproducible.

To make the workflow easier to handle, the code that generates the intermediate results should be placed in a separate script from the code that evaluates them and produces the final figures and tables.

\subsection{Reproducibility when original data cannot be shared}

In many cases, the data used for the analysis cannot be made available with the code and the reproducibility of the results based on the real data cannot be verified directly. For this purpose, synthetic data (pseudo data) can be generated that mimic the computational but not necessarily the statistical structure of the real data. This is entirely sufficient in this context, because the synthetic data are used solely for reproducibility checks---not for substantive inference. These synthetic data should be included in the code and data folder. Such synthetic data can be generated in R, for example, using the packages \texttt{synthpop} or \texttt{simdata}. 

For reproducibility, it is still important that the results generated using the real data are provided alongside the corresponding results based on the synthetic data, again using the same script with only the data object changed. Those interested in checking reproducibility can then attempt to reproduce the results obtained with the synthetic data. If this reproduction is successful, it is also likely that the corresponding results based on the real data would be reproducible if access to the real data were available.

\section{Existing reproducibility policies and perspectives for harmonised standards}

In this section, we first summarize the current reproducibility practices of selected journals. We then discuss why harmonised standards may be beneficial and outline conceptual dimensions that could inform the development of such standards. Here, standards refer to shared criteria for assessing reproducibility across journals and disciplines, not to uniform requirements regarding the level of achieved reproducibility.

\subsection{Current practices for selected journals with reproducibility policies}
\label{sec:journals}

Table~\ref{tab:reproducibility_policies} provides an overview of reproducibility practices in selected journals. These journals were selected as illustrative examples of established reproducibility policies. The list is not exhaustive but reflects a range of existing approaches across different fields. As noted in the introduction, several members of our author team are involved in the reproducibility processes of these journals, which provides additional practical insight into their procedures.

Although the approaches of these journals differ, all of them have implemented concrete measures to enhance computational reproducibility. These measures span a spectrum from lower-burden policies that primarily request that the analysis is described in sufficient detail, to more demanding procedures in which designated editors or reviewers actively check the reproducibility of all presented results.

\begin{table}[htp!]

\centering
\caption{Comparative overview of journal reproducibility policies}
\label{tab:reproducibility_policies}
\scriptsize
\begin{tabular}{
    >{\centering\arraybackslash}p{3cm}
    >{\centering\arraybackslash}p{1.5cm}
    >{\centering\arraybackslash}p{1.5cm}
    >{\centering\arraybackslash}p{1.5cm}
    >{\centering\arraybackslash}p{2cm}
    >{\centering\arraybackslash}p{1.5cm}
}
\centering{\textbf{Journal}} & \textbf{Code requirement} & \textbf{Data requirement} & \textbf{Independent check} & \textbf{Verification through} & \textbf{Timing} \\
\hline
Annals of Internal Medicine & encouraged & methods description only & no & not applicable & submission \\[20pt]
Biometrical Journal & yes & open or synthetic & yes & reproducible research editors & pre-final acceptance  \\[20pt]
The BMJ & yes & repository preferred & no & editors/readers (scrutiny) & submission \\[20pt]
Computo & yes (notebooks/CI) & open (or mock version if protected) & yes & Github Actions & submission \\[20pt]
Journal of Computational and Graphical Statistics & yes & repository encouraged & no & editors/reviewers & submission \\[20pt]
Journal of Statistical Software & yes (software focus) & open & yes & reviewers \& replication editors & first submission \& pre-final acceptance \\[20pt]
Journal of the American Statistical Association & yes (checklist) & data info provided & yes & associate editors for reproducibility & revision \\[20pt]
Meta-Psychology & yes & open (or justified) & yes & check team (stat.\ editor) & post-acceptance (conditional) \\[20pt]
Technometrics & yes (partial) & not applicable & no & editors & revision \\
\hline
\end{tabular}
\vspace{0.5em}

\begin{minipage}{\linewidth}
\scriptsize
\textbf{Journal}: The name of the academic publication whose reproducibility policy is being reviewed.

\textbf{Code requirement}: Indicates whether the journal explicitly mandates that authors submit the analytical code used in the study. The requirements can range from a simple \lq\lq yes'' to specific conditions like requiring a \lq\lq checklist'' or focusing on \lq\lq software''.

\textbf{Data requirement}: The policy regarding the provision of the underlying data. This spectrum includes requiring data to be \lq\lq open'' or that at least \lq\lq synthetic'' data should be provided, or mandating that a \lq\lq methods description only'' for data access is provided.

\textbf{Independent check}: Whether the journal performs an active, execution-based review by a third party (not the original author or initial reviewer) to confirm that the results can be computationally reproduced.

\textbf{Verification through}: Specifies the designated individual or group responsible for reviewing, checking, or scrutinizing the submitted reproducibility materials, such as \lq\lq reproducible research editors'', \lq\lq associate editors'', or standard \lq\lq editors/reviewers''.

\textbf{Timing}: The specific stage during the manuscript process when the submission of reproducibility materials or the verification check occurs, such as at \lq\lq submission'', during the \lq\lq revision stage'', or \lq\lq post-acceptance''.
\end{minipage}
\end{table}

The Annals of Internal Medicine requests that methods be described with sufficient detail to enable reproducibility and additionally advocates the inclusion of well-annotated source code.

The Biometrical Journal primarily targets methodological statisticians, most of whom use R. Manuscripts undergo the usual peer-review process and can be rejected or conditionally accepted. After conditional acceptance, a reproducibility review process is conducted by a dedicated team of four Reproducible Research Editors. Authors are advised to follow a list of guidelines to satisfy the journal's reproducibility requirements. There are different categories of reproducibility (e.g., fully reproducible and various categories of partially reproducible).

The BMJ, Journal of Computational and Graphical Statistics, and Technometrics have introduced a requirement for authors to submit the analysis code for all research articles. The code must be provided as a supplementary file and, preferably, additionally deposited in an open repository such as GitHub, GitLab, Open Science Framework, or Zenodo. While they do not conduct formal reproducibility reviews, the submitted code should be machine-readable, clearly annotated, and runnable with standard statistical software. Editors and readers may inspect the code if needed, and authors are expected to document software versions and relevant packages. Interestingly, Technometrics only require reproducibility of one table or figure.

Computo requires numerical reproducibility through an integrated workflow based on executable notebooks, virtual environments, and continuous integration. Unlike traditional journals, the article in Computo is generated directly from the underlying code and environment, making the computational workflow itself an important component of the contribution. When parts of the analysis must be run externally, authors must supply both the intermediate results and the code needed to import them into the notebook. All code required to generate the results, including code used for externally run steps, must be provided, and is assessed during peer review and/or editorial checks.

The Journal of Statistical Software sets full reproducibility as a requirement for submissions: manuscripts without complete reproducibility are often returned to authors prior to review.  Additional reproducibility checks are completed after review and manuscripts remain conditionally accepted until all requirements are met. This includes figures, tables, and any other outputs reported in the manuscript. A distinctive feature of the journal is its focus on software packages; empirical analyses are mostly illustrative only. The general guideline is that a reader with sufficient background knowledge should be able to reproduce the results in a reasonable amount of time. The software must be available in external repositories but the version at publication is stored with the manuscript.

For the Journal of the American Statistical Association, authors are asked to provide reproducibility materials (data, code, workflow) if the manuscript is invited for revision \cite{wrobel2024partnering}. Reproducibility is reviewed by Associate Editors for Reproducibility, of whom the journal currently has sixteen. The journal provides an Author Contributions Checklist that authors must complete (available as an R Markdown or a Word document). An example folder structure illustrating a possible organisation of research materials is also provided in a template GitHub repository. The checklist is divided into three parts that must be addressed in detail: (i) information on the data, (ii) details of the software and code, and (iii) questions on the reproducibility workflow.

Meta-Psychology requires conditionally accepted articles to be fully reproducible. To verify reproducibility, an independent check is conducted by one of the ten members of the reproducibility team with different areas of expertise. Submissions must include open data, well-documented code, and research materials, typically provided via an Open Science Framework project page; exceptions are granted only in justified cases. Reproducible submissions may receive an Open Science Badge, and authors are expected to cooperate with replicators or readers raising methodological questions even after publication. For more complex analyses, the use of containers to share computational environments is encouraged.

The journals discussed above represent only a subset of existing initiatives. Other journals that currently have appointed dedicated reproducibility editors include American Antiquity, Information Systems, the INFORMS Journal on Data Science, the Journal of Archaeological Science, the Journal of Economics Surveys, Political Analysis, and Research Synthesis Methods. These journals likewise implement various policies aimed at strengthening the reproducibility of empirical research.

\subsection{Towards harmonised reproducibility standards}

As seen in the previous subsection, journals currently provide very heterogeneous guidance on how authors should prepare reproducible submissions. These guidelines range from brief, high-level suggestions to extensive lists of required tasks or detailed templates that must be completed. While more elaborate reproducibility policies can facilitate a structured assessment, they may also discourage potential authors---especially when requirements are time-consuming or when policies impose constraints that are difficult to meet (e.g., mandatory public data availability). This heterogeneity makes it difficult for authors to navigate expectations and lengthens the time to resubmit papers to alternative journals following rejections.  Furthermore, it complicates how journals communicate their reproducibility standards to readers and stakeholders.

To address this challenge, we suggest considering the development of a harmonised, multi-tier system of reproducibility standards. Such a system would distinguish different aspects and levels of reproducibility. Importantly, these levels would not necessarily reflect differences in the amount of work required but rather in the aspects of reproducibility that are achieved. A graded scheme would acknowledge that reproducibility is multifaceted and enable consistent assessment across studies and disciplines (e.g. technical code review templates of Hanselman \cite{Hanselman:2025} used for conferences). In the subsection that follows this one, we tentatively outline several ideas that could serve as a basis for developing such a multi-tier system of reproducibility standards. Systems of standards already exist in other domains---for example, for assessing FAIR data (e.g. \url{https://www.f-uji.net/index.php?action=methods}) or for peer review of statistical software (e.g. \url{https://stats-devguide.ropensci.org}).

Quantifying the levels of reproducibility could offer several benefits. Currently, scientific output is primarily evaluated through publication counts, journal impact factors, and citation-based metrics such as the h-index. Despite increasing recognition of reproducibility as a scientific value, it is rarely incorporated into commonly used indicators. A harmonised standard could, in principle, lay the groundwork for reproducibility-aware evaluation metrics or indices. Such metrics could incorporate not only different reproducibility levels but also established dimensions such as the FAIR principles, thereby providing a more comprehensive representation of research transparency. Their calibration would reflect the priorities of the scientific community and could support a more nuanced assessment of research quality.

Many journals label articles as \lq\lq reproducible'', but this designation can represent very different aspects, leading to inconsistent signals for readers. A clear and structured system of standards would make transparent which aspects of reproducibility were considered, omitted, or feasible in a given context. Such standards need not imply uniform requirements across disciplines; rather, they would offer a shared framework for describing reproducibility in a consistent way. This would help journals communicate their reproducibility aims and scope more effectively and clearly for the benefit of publishers and editorial boards. Clearer standards would also improve communication with stakeholders, including funding agencies, data repositories, and research institutions, which increasingly emphasize transparency and reproducibility.

In addition, a graded structure can help journals align their reproducibility policies with their available resources. For example, journals might initially focus on clear availability requirements and basic checks of submitted materials and then, where feasible, move towards more thorough reproducibility review, potentially supported by dedicated editors or reviewers.

Authors, too, could profit from such a standardised and potentially quantifiable system. A well-defined multi-tier structure would allow authors to quicker and better assess the level of reproducibility they can reasonably achieve and demonstrate that their work meets or exceeds a clearly defined threshold. This could strengthen the perceived value of reproducible research in academic careers while providing a structured but non-prescriptive framework. Additionally, when needing to submit the manuscript to another journal, a harmonised system would reduce the burden to substantially restructure the code and data supplement when submitting the manuscript to another journal.

Overall, while we introduce initial ideas rather than finalized proposals, we believe that harmonised reproducibility standards could support clearer communication, reduce unnecessary burden on authors and reviewers, and provide a foundation for future developments in how reproducibility is evaluated across scientific fields.

\subsection{Conceptual dimensions for structuring reproducibility standards}

A multi-tier system of reproducibility standards requires a clear understanding of which aspects of reproducibility should meaningfully be differentiated, assessed and weighted with respect to their importance. While we do not propose a concrete implementation, we outline several conceptual dimensions that could, in principle, form the basis of such a system.

It is our hope that the following dimensions fuel future discussions about how reproducibility can be evaluated in a transparent and systematic way.

\begin{itemize}
\item \textbf{(A) Availability of materials.} This dimension captures which components of the reproducible research workflow are made available. Relevant elements include code, data, metadata, and descriptions of the computational environment. Different levels could distinguish, for example, between code-only availability, code-and-data availability, or additionally the provision of structured metadata aligned with the FAIR principles.

\item \textbf{(B) Verification scope.} Although only an explicit check of computational reproducibility can definitively confirm that the results are reproducible, such verification can be resource-intensive and not always feasible. More lightweight assessments can nevertheless provide meaningful indications of how likely reproducibility is to hold. This dimension could therefore distinguish the depth of verification applied to the code and data supplement, with levels increasing in rigour, for example: (i) checking the completeness of the materials, (ii) assessing their quality and ease of use, and (iii) verifying computational reproducibility of some or all results.

\item \textbf{(C) Verification source.} Depending on whether and if so, by whom the materials have been checked, different degrees of trust can be placed in the reproducibility of these checks. The lowest degree of trust applies when no checks have been performed. Higher trust applies when the authors state that they have thoroughly reviewed the materials themselves prior to submission. This might be further distinguished by whether the code author or someone else on the same research team conducted the checks. The latter would be associated with a higher level of trust. The highest trust applies when the materials have been examined on the journal side, for example by designated reviewers or reproducibility editors.

\item \textbf{(D) Scope of reproducibility.} Studies differ in the extent to which their results can be reproduced. Some analyses allow full computational reproducibility, whereas others can only achieve partial reproducibility---for instance, when original data cannot be shared and synthetic data are provided. However, there are various other reasons why results may be only partially reproducible. For example, if an analysis is computationally extremely expensive and no reproducible intermediate results are available that allow reproducibility spot checks, the analysis may not be reproducible without substantial computational resources. Another reason may be that generated random numbers in a simulation study are not reproducible (e.g., due to issues with seed handling), in which case the results will not be exactly reproducible; still, the substantive conclusions from the analysis may remain valid if a large enough number of replications has been used. A tiered system could distinguish such cases in a principled manner.

\item \textbf{(E) Code quality.} The quality of the code directly affects how easily the results can be reproduced. Our instructions in Section~\ref{sec:guidelines} could provide a basis for metrics for this dimension. Important points include, for example, the clarity of the code and data supplement, the extent of commenting, and the availability of intermediate results. In principle, these aspects could be weighted differently and aggregated into a fine-grained scoring system. However, developing such a system would require careful calibration and could therefore be more of a longer-term goal. A pragmatic first step would be to define a few quality levels that capture meaningful differences while remaining easy to apply.
\end{itemize}

These dimensions are not exhaustive and offer a conceptual foundation for thinking about how reproducibility standards could be organised in a multi-tier framework that incorporates various aspects and degrees of reproducibility. We invite the community to join us in refining and expanding these ideas to make them more operable and thus adoptable by journals and publishers.

\section{Discussion}

\subsection{Challenges in operationalising a multi-tier system of reproducibility standards}

 To make our concept of a multi-tier system for measuring the degree of reproducibility in empirical research operational, various refinements and specifications will be required. For example, it must be clarified how the different dimensions should be represented---whether in the form of a single aggregated metric or as a multidimensional profile---and how different aspects should be weighted. Given that some dimensions are interdependent or not assessable in all cases, such choices require careful consideration. In our view, however, it is not necessary for an initial operational system to be flawless. Even a pragmatic first version could play an important role in increasing the visibility of reproducibility practices and thereby further raising awareness of their importance.

As also noted in the previous section, our instructions from Section~\ref{sec:guidelines} could serve as an initial concrete basis for the fifth proposed dimension of reproducibility, \lq\lq code quality''. These instructions draw on extensive experience with code review and are therefore well suited as a foundation for generally applicable guidelines.

\subsection{The importance of journal policies involving reproducibility checking}

As we mentioned, the thoroughness of reproducibility policies lies on a spectrum. One of the most far-reaching and effective approaches is for the journal itself to check reproducibility. However, this requires the involvement of reviewers or dedicated reproducible research editors and entails not only substantial coordination but also considerable effort. For many journals, especially those with limited resources, such procedures may currently be difficult to implement in practice. Nevertheless, the proportion of journals pursuing such policies should increase in order to raise the share of reliably reproducible results in practice. For scientific articles, it is generally accepted that journals without peer review necessarily produce output of lower quality, which is why peer review of scientific articles is recognised as an essential component of quality assurance in research. Because code and data supplements form the crucial basis for the corresponding scientific articles, it appears no less important to subject these supplements to a review process as well---this can uncover errors in analyses and thus would represent a concrete and constructive step toward addressing the replicability crisis.

If reproducibility were a requirement, that could also contribute to earlier detection of errors. At present, journals that conduct reproducibility checks typically require authors to submit their code and data supplements only after the article has been conditionally accepted on the basis of its scientific content. One reason for this practice is the desire to keep the reproducibility review workload manageable. Another plausible reason is that authors may be less willing to prepare a clean code and data supplement that enables reproducibility when they do not yet know whether the article will be accepted by the target journal. This reduced willingness is largely driven by the fact that most journals do not require reproducibility at all.

If, however, more journals were to require the code and data supplement already at initial submission, it would likely become easier for journals to make such requests. Once the submission of a code and data supplement becomes a common expectation across journals, authors can reasonably expect that they will need to provide it at the next journal as well if the paper is rejected. Submitting the code and data supplement at the initial stage can benefit both journals and authors, as it reduces the risk that a paper must be rejected shortly before publication because its key results turn out to not be reproducible due to errors in the code. Such cases are not the norm, but they do occur in practice, according to our editorial experience.

It is important to distinguish whether the data used can be made publicly available or whether legal, ethical, or proprietary constraints prevent public release. In the latter cases, complete reproducibility for the scientific community is not possible, but, if feasible, the data should at least be made available to reviewers or reproducibility editors to enable independent verification of the results. Authors should also discuss whether certain conditions allow other researchers to access the data for further research.

\subsection{The role of reviewers}

In this paper, we have focused on the perspective of authors and journals because we believe that these scientific actors have the most immediate impact on the dissemination of reproducibility practices. However, as indicated above, reviewers also have an important influence on ensuring reproducibility, albeit, often with fewer incentives. This does not diminish the fact that reviewers should care about whether the results they assess are correct and reproducible; overlooking this aspect risks leaving their evaluation incomplete. 

Even without performing time-consuming reproducibility checks, reviewers can already make valuable contributions. For example, it would be helpful if they pointed out when code and data are missing or, one step further, whether code is available to reproduce all figures and tables in the paper. If a greater level of commitment is to be expected from reviewers, clear guidelines and easily verifiable quality standards could be highly beneficial. 

In addition, incentives for reviewers could be strengthened by acknowledging their verification of reproducibility through publicly visible information, as is already common practice in the peer review of scientific articles. For instance, Meta-Psychology already acknowledges the reproducibility checkers alongside the peer reviewers, which may serve as a model for other journals. A universal advantage for reviewers is that examining the code of others in their field may allow them to learn new coding practices that they can apply in their own research. Moreover, in some cases, reviewing the code can provide deeper insights into the methodology discussed in the article.

While reproducing the reported results is central to computational reproducibility, it does not reveal whether the description of the analysis actually matches the implementation within the code. This is an aspect that is often overlooked and where peer reviewers may be able to provide valuable insight in the future.

\subsection{Future perspectives: education and automation}

Authors, journals, and code reviewers could all benefit from clear training materials that illustrate the various aspects of conducting reproducible research. These could take the form of interactive tutorials or short videos. However, care should be taken to ensure clarity and conciseness, in order to avoid overwhelming the intended users of such materials, as evidenced by reports on corresponding training courses \cite{Heise:2023}. Examples of such materials within the R community include the pedagogical resources developed by Jennifer Bryan and her collaborators, such as the \lq\lq What They Forgot to Teach You About R'' site (\url{https://rstats.wtf}) and the concise guidance on file naming (\url{https://github.com/jennybc/how-to-name-files}), which illustrate how concise, practical guidance can support authors in adopting cleaner workflows that support reproducibility.

In the long term, reproducibility checks may be increasingly automated. Ever-improving containerization (e.g. Docker images, MaRDI Packaging System \cite{MaPS:2024}), or environment management tools, such as \texttt{renv} for R or \texttt{conda} for Python, facilitate the description of the computational environment. Natural language-based models could assist reproducibility editors or reviewers by automating parts of the workflow, supporting communication with authors, or performing preliminary checks before human evaluation. However, even if reproducibility checks eventually become fully automated, attention should still be paid to the accessibility and readability of the code so that it remains interpretable for humans. This is particularly important because the code and data supplement is a core component of empirical research.

\section{Conclusions}
In this paper, our aim was to address factors that currently hinder many authors and journals from engaging more strongly with the reproducibility of data-based analyses, thereby hopefully contributing to a broader dissemination of reproducible research.

To this end, we first summarized the wide range of self-serving benefits of reproducibility for both authors and journals, with the goal of strengthening the motivation of these key actors. We then proposed pragmatic and easily implementable instructions for conducting reproducible analyses, based on our extensive experience with code review, to counteract the common feeling of being overwhelmed, often caused by overly complex guidelines. Finally, we suggested the development of a multi-tier system for evaluating reproducibility, incorporating several dimensions of reproducibility. Once developed in more detail, such a system could enable journals to report the level of reproducibility of published empirical studies in a standardised way, which may further promote the dissemination, institutionalisation, and standardisation of reproducibility while also simplifying expectations for both journals and authors.

A central foundation of this system is the recognition that reproducibility exists on a continuum. Therefore, even small, pragmatic steps by authors, journals, and reviewers can meaningfully contribute to improving the quality assurance of empirical research.

\section*{Acknowledgements}
We gratefully acknowledge that the analogy used in the introduction, which compares the role of code and data in empirical research to the role of proofs in pure mathematics, was originally suggested by Dr.\ Fabian Scheipl in an unrelated prior discussion.

RH, LN, and OZ acknowledge funding from the Deutsche Forschungsgemeinschaft (DFG, German Research Foundation) under the National Research Data Infrastructure (NFDI) within the Mathematical Research Data Initiative (MaRDI), project number 460135501, NFDI 29/1. RK acknowledges partial funding of this work from the UK Engineering and Physical Sciences Research Council under grant EP/T021020/1. ALB acknowledges funding from the DFG under projects BO3139/9-1, 7-2 and 10-1.

\section*{Competing Interests}
For transparency, the following roles are disclosed. RH, TU, MK, and BH serve as Reproducible Research Editors for the Biometrical Journal. RK serves as Co-Editor-in-Chief for the Journal of Statistical Software and Journal of Computational and Graphical Statistics, Associate Editor for Technometrics, and a member of the ROpenSci Statistical Software Peer Review Standards Board.  CJP and JW serve as Associate Editors of Reproducibility at the Journal of the American Statistical Association. JC serves as the Editor-in-Chief of Computo and PN serves as Associate Editor of Computo; both founded the journal. MH serves as Reproducibility and Open Science Transfer Coordinator of the Munich Center of Machine Learning (MCML). LB serves as Reproducibility Coordinator and RC serves as Editor-in-Chief at Meta-Psychology.

\section*{{S}upplementary material}
The supplementary material accompanying this article is available on the arXiv page.

\bibliographystyle{unsrtnat}
\bibliography{bibliography_paper}

\end{document}